\documentclass[twocolumn,aps,pre,showpacs,amsmath,amsfonts,amssymb,floatfix,pre]{revtex4}
\usepackage{graphicx}
\usepackage{dcolumn}
\usepackage{bm}
\usepackage[bookmarks=false]{hyperref}

\begin{document}

\title{Spectral Properties of Directed Random Networks with Modular Structure }

\author{Sarika Jalan}
\email{sarikajalan9@gmail.com}
\affiliation{School of Sciences, Indian Institute of Technology Indore, IET-DAVV Campus Khandwa Road, Indore 452017, India}
\affiliation{Department of Physics and Centre for Computational Science and Engineering, National University of Singapore, 117456, Republic of Singapore}

\author{Guimei Zhu}
\email{zhugm07@gmail.com}
\affiliation{NUS Graduate School for Integrative Sciences and Engineering, 117456, Republic of Singapore}
\affiliation{Department of Physics and Centre for Computational Science and Engineering, 
National University of Singapore, 117546, Republic of Singapore}

\author{ Baowen Li}
\email{phylibw@nus.edu.sg}
\affiliation{NUS Graduate School for Integrative Sciences and Engineering, 117456, Republic of Singapore}
\affiliation{Department of Physics and Centre for Computational Science and Engineering, 
National University of Singapore, 117546, Republic of Singapore}

\begin{abstract}
We study spectra of directed networks with inhibitory and excitatory couplings. 
We investigate in particular eigenvector localization 
properties of various model networks for different value of correlation among their entries. 
Spectra of random networks, with completely uncorrelated entries  
show a circular distribution with delocalized eigenvectors, where as networks with
correlated entries have localized eigenvectors. In order to understand the origin of 
localization we track the spectra as a function of connection probability and directionality.
As connections are made directed, eigenstates start occurring in complex conjugate pairs 
and the eigenvalue distribution combined with the localization measure shows a
 rich pattern.
Moreover, for a very well distinguished community structure, the whole spectrum is localized 
except few eigenstates at boundary of the circular distribution. 
As the network deviates from the community structure there is a sudden change in the localization
property for a very small value of deformation from the perfect community structure. We search
for this effect for the whole range of correlation strengths and for different community
configurations. 
Furthermore, we investigate spectral properties of a metabolic network of zebrafish, and 
compare them with those of the model networks.

\end{abstract}
\pacs{89.75.Hc,89.90.+n}

\maketitle

\section{Introduction}

Statistical properties of large random matrices have been investigated intensively since 
the early 
1950s and have turned out to be valuable tools for both qualitative and quantitative descriptions of 
spectral properties of complex systems \cite{mehta}. Examples include, complicated interactions in 
nuclei, quantum mechanical aspects of chaos, and the stock market. Recent success has been achieved in systems having an 
underlying network structure \cite{rev-rmt}. In applications of random matrix theory to physical 
problems, it is generally assumed that the details of the physical systems are less important for many 
statistical properties of interest. Often it turns out that important statistical properties such as 
the distribution of eigenvalues or the spacings of energy levels in quantum systems are well described by 
the respective properties of random matrices that respect the same symmetries as the physical system 
\cite{Hakke}.

Recent investigations of statistical properties of eigenvalues of networks show that bulk  
eigenvalue spacings follow Gaussian orthogonal ensemble (GOE) predictions of random matrix theory (RMT) 
\cite{Vattay-NJP,pre2007a,pre2007b}, whereas the eigenvalue distribution itself depend on network 
properties \cite{spectra-net,Aguiar-SF,pre2007a,pre2007b,pre2010a}. These analyses show that the spectral 
distributions of bulk eigenvalues of random networks follow Wigner's semi circular distribution, for 
scalefree networks \cite{BA} the bulk eigenvalues follow a triangular distribution, for smallworld 
networks \cite{SW} they form a multi peak distribution, and for realworld networks they show 
one of the above mentioned distributions with a high peak at zero eigenvalues \cite{pre2007a,pre2010a}. Further 
investigations of the community structure in networks show the transition to Wigner's semi circular 
distribution as the community structure is destroyed. Even a very small deformation from the community structure 
gets captured by the eigenvalues distribution \cite{pre2009b}.

All the above investigations have been performed for undirected networks leading to 
a symmetric adjacency 
matrix and for which eigenvalues are real. Many complex systems are described by directed networks 
\cite{rev-network}, for example, the World Wide Web, neural networks, protein interaction networks and many 
social networks are directed or asymmetrically weighted, and hence lead to complex eigenvalues of the 
corresponding adjacency matrices. One of the most important features shown by realworld networks 
is the existence of modular or community structures \cite{Newman,Amaral,community}. The study of 
community structures helps to elucidate the organization of networks, and eventually could be related 
to the functionality of groups of nodes. Regardless of the type of realworld networks in terms of the 
degree and other structural properties \cite{rev-network}, it is possible to distinguish communities throughout the whole networks \cite{Newman}. However, the question of the definition of the community is 
problematic. Usually community is assigned to the nodes that are connected densely among 
themselves, and are only sparsely connected with other nodes outside the community.
 
So far most of the theoretical investigations and applications of RMT have focused on symmetric 
matrices. Asymmetric matrices or non-Hermitian matrices are less well understood. A classic result for 
non-Hermitian matrices in random matrix theory is 
Girko's circle law \cite{Girko} which states that, for large $N$, the eigenvalues of an $N \times N$ 
asymmetric matrix lie uniformly within the unit circle in the complex plane, if the elements are 
chosen from a distribution with zero mean and variance $1/N$. When partial symmetry is included, the 
circle changes to an ellipse \cite{density-ellipse}. Recently, numerical studies on spectral 
properties of asymmetric matrices have been done in the context of Google networks 
\cite{google-net,google-net2}. Additionally the spectral density of non-Hermitian sparse matrices has 
been analyzed in the Ref. \cite{directed-sparse-RMT}.

In this paper we evaluate the applicability of RMT for the distribution of eigenvalues of directed 
networks where entries in the corresponding matrix take values motivated by inhibitory and 
excitatory coupling between nodes. In particular, we investigate spectral properties of directed 
networks having a community structure. The matrix corresponding to these networks is different from the 
non-Hermitian random matrices studied under a RMT framework. The main differences are that (a) the matrix is not 
{\it random}, i.e. the position of entries in the matrix depends on the underlying network structure; (b) real 
world networks and model networks are sparse and consequently most of the entries in the corresponding 
matrix are zero; and (c) most importantly, the entries in the matrix take values $0$, $-1$ and $1$. 
These entries are motivated by the excitatory and inhibitory synapse between neurons leading to positive and 
negative values of the connections between them.

The spectra of matrices having entries $0$, $1$ and $-1$, known as Seidel 
matrices, have been investigated intensively 
in the literature. In graph theory, the Seidel adjacency matrix of a simple graph G is a symmetric matrix
having $0$ at the diagonal entries,  $G_{ij}= -1$ if node $i$ is
neighbor of node $j$, and other entries of this matrix taking the value $1$. For Seidel matrices
$G_{ij}=G_{ji}$. The multiset of eigenvalues of this matrix is called the Seidel spectrum. The
Seidel matrix was introduced by van Lint and Seidel \cite{Seidel_1966} and extensively exploited by 
Seidel and co-workers. Seidel matrix provides insight to the properties of 
graphs, thus 
leading to several applications such as modeling social situations, spin glasses in physics, and 
clustering of data \cite{Seidel_regular}. 

Note that, except for the matrix entries being binary
$1, -1$ and $0$, Seidel matrices are completely different from the matrix representation considered in the present paper.
The first difference between Seidel matrices and the matrices investigated 
herein is that the latter are directed (asymmetric) matrices, i.e., $A_{ij} \ne A_{ji}$
depending upon the value of $\tau$, a measure of directionality or correlation. Consequently, we get 
complex eigenvalues following the universal
Girko's law for the case when the matrix is completely uncorrelated. The second most important
difference is that we have $A_{ij}=0$ for the nodes that are not connected, 
i.e., vertices that
are not neighbors, whereas Seidel matrices have the entries $1$ for non neighboring nodes.
 This second property of our matrices again leads to a very crucial
difference called 'sparseness', which affects the spectral properties to a great extent
in comparison to the matrices corresponding to fully connected networks.
The third important difference between the matrix we have considered in this paper and the Seidel
matrices we found in the literature is the underlying structure: Seidel matrices are mostly
investigated for regular graphs (such as strong regular graphs and two-graphs) \cite{Seidel_Spectra_notes}, whereas the
matrices that we are investigating in the paper are {\it random with some structure}, 
corresponding to ER graphs,
BA scalefree graphs, modular graphs and modular realworld networks. There are cases        
in the literature where the Seidel matrix of ER graphs and realworld graphs have
been investigated; in the following we provide details of these studies only, as they seem to be close
to our studies at least in terms of the structure of the underlying network (
Seidel matrices of random graphs have been studied mainly for the graph bisection problem.) Reference 
\cite{Seidel_random_bisection} shows through numerical analysis that Seidel representations produce the
smallest cuts for random ER graphs than other representations such as the 
adjacency and Laplacian matrices.

Spectral densities of directed random networks with entries $0$, $1$ and $-1$ have also been studied 
analytically and numerically in Ref. \cite{Abbott-prl2006}. In Ref. \cite{Abbott-prl2006} eigenvalues spectra of large 
random matrices with excitatory and inhibitory columns drawn from a distribution with different means 
and equal or different variances have been computed. Interestingly, the spectra of these random matrices 
(with $0$ and $1$, $-1$ entries) also follow Girko's circle law.

The main goal of the present work is to investigate the changes in the spectral properties of 
the corresponding matrix as the underlying directed network deviates from the perfect community structure. Furthermore, in order to understand the origin of delocalized eigenvectors,
we track the spectra as directionality 
is introduced into the networks, i.e. connections are made directed with probability $\tau$. 
Apart from the :spectral distribution we study eigenvectors localization properties of model networks and a 
realworld networks.

\section{Model}
\label{m_Rand}
We generate networks with a community structure as follows. First, $m$ Erd\"os-R\'enyi 
random networks \cite{ER} with connection probability $p$ are constructed; the spectral behavior of the matrix 
corresponding to each of these sub-networks (blocks) separately follows GOE statistics 
\cite{pre2009b}. The matrix corresponding to the full network would be an $m$ block diagonal matrix. 
Then, with probability $p_r$ connections are rewired as follows. A pair of 
nodes from a sub-network is chosen randomly. Subnetwork is also chosen with equal probability. The connection 
between these two nodes is deleted, and a new connection between one of the node in the chosen pair
and a randomly chosen node outside to this sub-network is introduced. 
This configuration leads to an $m$ block matrix with, blocks having entries $1$ with probability $p$, 
and off-diagonal blocks having the entries $1$ with probability $p_r$. The above networks can be casted 
in the following form:
\begin{equation}
A = A_0 + A_{p_r}
\label{net}
\end{equation}

where $A_0$ is the direct sum of all the blocks ($A_0 = A_1 \oplus A_2 \oplus \hdots A_m $). 
The directionality of the edges is introduced as follows. A node is assigned
equal probability of being 
inhibitory or excitatory. All edges starting from an excitatory node would take value $1$, and those 
starting from an inhibitory node would take the value $-1$. This means that 
the corresponding adjacency matrix 
$A$ has entries as follows: if the $ith$ node is excitatory, $A_{ij} = 1$ for all the $j$ nodes that have 
connections with $i$, and if the $ith$ node is inhibitory, $A_{ij}=-1$. The corresponding matrix 
has following two properties:
(a) $|A_{ij}|  = |A_{ji}|$, and (b) a row has all entries either $1$ or $-1$. This configuration 
would lead to zero mean ($\sum_{i,j}A_{i,j}\sim 0$), and standard deviation $ \sigma^2 \sim \tau$. 
We denote the eigenvalues of network by 
$\lambda_n,\,\,n=1,\dots,m N$, where $N$ is the size of sub-network, and $m$ is the number of 
sub-networks. Note that the size of each sub-network may be different, but for simplicity we consider 
equal size here. 

Connections are rewired with the probability $0 < p_r < 1$, where $p_r$ defines the ratio of 
inter-community ($N_{inter}$) to intra-community ($N_{intra}$) connections. Starting with two ER 
sub-networks of dimension $N$, a fraction $p_r$ of the connections is rewired. The average degree of  
network for each $p_r$ remains the same as that of the initial configuration. Nodes are inhibitory or excitatory with 
equal probability, which implies an approximately equal number of $1$ and $-1$ entries randomly 
distributed in the corresponding matrix. For $p_r \sim 0.5$, the whole structure becomes a directed random  
network, and for $p_r \sim 1$ which indicates that all the connections are rewired, the network takes the 
form where nodes in $A_1 (A_2)$ are connected to those in $A_2 (A_1)$.

\section{Spectra}
In the following we present the results for $N=1000$, $m=2$, and $p=0.02$. The average degree of the 
network can be calculated as $<k> \sim pN = 20$. The eigenvalues spectra of 
$A_1$ and $A_2$ for $p_r=0$ follow Girko's law of non-Hermitian matrices \cite{Girko}. We numerically 
diagonalize the adjacency matrix of the network to obtain a set of eigenvalues $\lambda_k = R_k + i 
I_k$, and eigenvector $\psi_{i}$. To characterize the localization properties of the eigenvectors $\psi_i$, we 
use the inverse participation ratio (IPR) defined by,
\begin{equation}
I_i = \frac{(\sum_{j} |\psi_i(j)|^2)^2}{\sum_j |\psi_i(j)|^4}
\end{equation}
where $\psi_i(j), j=1, \hdots , N$ are the components of the eigenvector $\psi^i$. The meaning of $I$ is 
illustrated by two limiting cases : (i) $I_i =N$ for a vector with identical components $\psi_i(j) \equiv 
1/\sqrt{N}$ whereas (ii) $I_i = 1$ for  a vector with one component $\psi_i(j)=1$ and the remainders 
zero. Thus, the IPR gives the effective number of nodes on which an eigenstate is 
localized. 

\begin{figure}
\includegraphics[width=0.85\columnwidth,height=5.6cm]{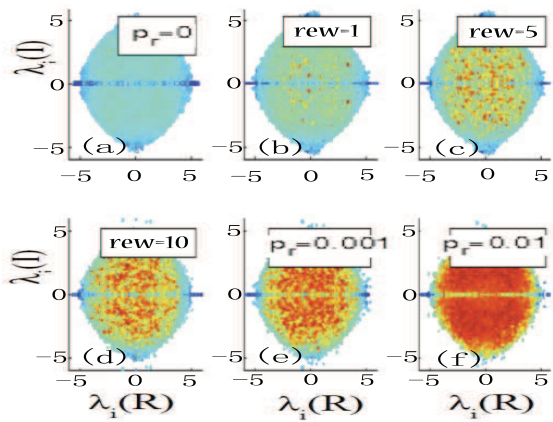}
\caption{(Color online) Spectra with IPR for networks starting with two communities of size $N=1000$
 and then rewiring connections between communities with different probabilities. 
Average degree is $<k>=20$. The real and imaginary parts of the eigenvalues
are plotted on the x and y axes, respectively.
(b) One and (c) five rewirings are plotted
for an ensemble average of 100. For other rewiring probabilities the
spectra are plotted for an ensemble average of 10. The rewiring probability
is denoted by $p_r$. The value of the IPR varies from (a)-(f) $I=0$ [
dark gray, blue, patches 
at the corners] to (b)-(f) $I=1100$ [dark gray, red, points inside the 
circular region]. Light gray regions inside the circle in (a) - (c) depict less delocalization than the
dark gray patches at the corners. Light gray area inside the circular regions and the horizontal
line at x axis of (d)- (e)
are even less delocalized than the peripheral light gray areas that have the 
same level of delocalization as the interior of (a)-(c). }
\label{fig_m_Rand}
\end{figure}

Figure~(\ref{fig_m_Rand}) plots eigenvalues together with the IPR for the model described in the section 
\ref{m_Rand}. Subfigures correspond to different rewiring probabilities. For each rewiring 
probability eigenvalues are distributed homogeneously in a circular region. Gray shading (color) 
denotes the value 
of IPR for the corresponding eigenstate. For Figure~(\ref{fig_m_Rand} a) with $p_r\sim 0$, which corresponds to the 
network having two 
communities connected with one single connection, large number of eigenstates 
have approximately equal IPR value close to the bottom of the shading (coloring). Few eigenstate lying at the boundary 
of the circle and a few near the real axis have low IPR values, indicating high participation in the 
localization.

Figure~(\ref{fig_m_Rand}) plots the results upto rewiring probability $p=0.01$, as rewiring
is increased further the spectral distribution remains exactly the same as for $p=0.01$ case
except that the light gray (green) hazy circular part becomes bigger and denser indicating that more  
eigenvalues are delocalized. As rewiring keeps on increasing, leading to the destruction of the community 
structure more and more eigenvalues join the light gray (green) big circle, and for 
$p_r = 0.1$ 
except for a few eigenvalues lying at the end of the real and imaginary axes all are delocalized. 
Additionally, the eigenstate near the real axis has relatively lower IPR values. One eigenvalue 
lying at the extreme end of the real axis is the most localized.  
For a further increase in rewiring probability there is not much of a 
change in the localization properties of eigenstates. The value
$p_r=0.5$ corresponds to the same expected number of intra and inter-community connections. At this value the 
network is a complete random network. A further increase in $p_r$ takes the network toward 
a bipartite type of structure.

\subsection{Fine rewiring}
Figure~(\ref{fig_m_Rand}) shows that the localization properties of a network 
change completely even 
for a very small value of $p_r$, as small as $p_r=0.1$. After this value no visible changes are seen in the 
localization properties of the bulk of the spectra. To gain insight into the
localization to 
delocalization transition we track the spectra for each rewiring between the communities (Eq.~\ref{net}). 
First four subfigures (Figure~\ref{fig_m_Rand} a - \ref{fig_m_Rand} d) plot the spectra along with the 
localization value for a very small number of rewiring All figures are plotted for an ensemble
over 100 random initial conditions for the networks. Figure~(\ref{fig_m_Rand} a) plots the 
spectra and IPR value for 
a perfect community network without any connections between them. This configuration leads to two 

disconnected components. With only one rewiring between two communities, we already start 
observing the delocalization of few eigenstates (Figure~\ref{fig_m_Rand} b). 
Each additional rewiring between communities leads to the delocalization of more eigenstates. 
For as few as 10 rewiring to the initial configuration (Figure~\ref{fig_m_Rand} a) a
large number of eigenstates are delocalized (Figure~\ref{fig_m_Rand} d). 
One can notice in Figure~(\ref{fig_m_Rand}) that delocalized eigenstates  
form a hazy ring around center \cite{Abbott-prl2006}. Eigenstates lying at the peripheral of the 
circle and those corresponding to the real axis delocalized in last for very large value of rewiring
probability between the communities.
For $p_r =0.01$, eigenstates near real axis and with large absolute
values are those which still maintain localization. As value of $p_r$
increases further except few eigenstates having largest
absolute values lying at real and imaginary axis are delocalized. Bulk
middle part delocalized completely. As value of $p_r$ increases further, bulk
middle part remains same, only eigenstates with largest absolute eigenvalues
at imaginary axis are more delocalized. For $p_r \sim 0.9$, the spectra remains similar to that of value $p_r=0.1$.

We would like to note that spectral properties of $m$ coupled $GOE$ random symmetric matrices have been 
investigated in details. They have been used to understand the behavior of quantum graphs \cite{Seba}.

\section{Comparison with Random Networks}
\label{Rand}
In order to gain insight into community induced spectral changes, we present results of random 
networks. In the following we first provide the eigenstate properties of random networks having 
different average degrees, and then track the spectral properties as 
parameters of $\tau$. 

\subsection{Completely uncorrelated (directed) network} 
First we consider 
completely uncorrelated random networks, for which $\tau=0$ and the 
connection probability is $p$.
Since there is  equal probability of a node
being inhibitory or excitatory, the expected number of $1$ and $-1$ entries is
approximately same. The mean and variance of this network can be calculated as $\mu=0$ and 
$\sigma^2=p$, respectively.

\begin{figure}
\includegraphics[width=0.8\columnwidth,height=5.5cm]{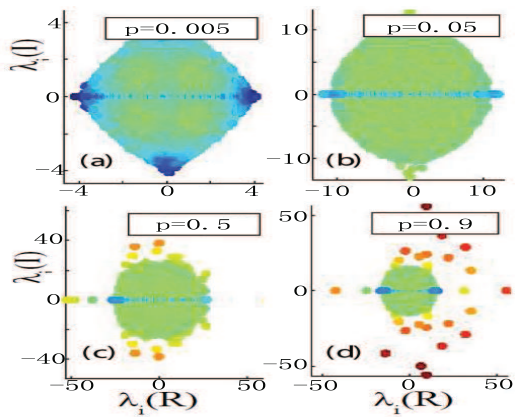}
\caption{(Color online) Spectra with the IPR for directed random networks having different connection
probabilities $p$.
The size of the networks $N=2000$, and spectra are plotted for 10 random 
realizations
for each value of $p$. The
real and imaginary parts of the eigenvalues are plotted on the x and y axes,
and gray shading (color) denotes the value of the corresponding
IPR. The value of the IPR varies from $I=0$ [dark gray, blue, patches in 
corners of (a)] 
to $I=2000$ [scattered dark gray, red, dots in (d)]. Light gray areas
 inside the circular
regions in (a)-(d) are less delocalized than the dark gray patches in (a). Light gray dots outside
the circular region in (c) are not as delocalized as the light gray regions described earlier.
Darker gray dots outside the circular region in (c) and (d)  are not as
 delocalized as the light 
gray dots in the same region and are  more delocalized than the dark gray dots in (d), which are most the localized. Note the
scale of the gray shading (color) axis for this case, which is different from other figures. (a)-(d) The horizontal line at x-axis shows the same level of 
localization as the
value of delocalization close to the most delocalized eigenstates.}
\label{fig_Rand}
\end{figure}

Figure~(\ref{fig_Rand}) plots eigenvalues together with the IPR for random networks with different
connections probabilities.

As we do not consider self connection, the diagonal elements of the matrix are zero, and hence 
eigenvalues are distributed around origin $0$. Eigenvalues are distributed in circular region of 
radius $\sqrt{N p (1-p)}$. For larger values of $p$, few pairs of eigenvalues get separated from the 
circular bulk.  The bulk circular region still lie between $-\sqrt{N p (1-p)}, \sqrt{N p (1-p)}$. For 
$p=1$ which corresponds to globally coupled network, there is one conjugate pair of eigenvalues with absolute 
value $\sqrt{N}$ which are nondegenerate, rest of eigenvalues are degenerate with values $1$ and 
$-1$.

Different connections probabilities lead to different expected average degree of the network. For low 
connection probability, $p=0.005$, one gets localized eigenstates at four corners of the real and 
imaginary axis.  Apart this four corners the eigenstates on real axis are also more localized. 
As connection probability 
increases overall distribution of eigenvalues remains same, i.e. homogeneous, except for larger $p$ 
few pairs of eigenvalues get separated from the bulk circular region \cite{google-net}. Except 
eigenstates lying on real axis, other eigenstates are toward dark gray (blue) denoting delocalization. 
The isolated eigenstates 
are maximal delocalized, which implies that eigenstates having large absolute eigenvalues are more 
random than the bulk part. Very small values of $p$ yields a delocalized spectra 
(Figure~\ref{fig_Rand} a), which may be due to the sparseness of connections.

Here we would like to mention that, the above behavior is for the
special arrangement of $1$ and $-1$ entries in the matrix, a row has all $1$ or $-1$ entries
depending upon whether the corresponding node is excitatory or inhibitory.
 For a random network with entries $1$ and $-1$ randomly distributed, the radius of circular region 
scales with average degree of the network, i.e. $\sqrt{p N}$, and all eigenvalues lie within the bulk 
regions even for larger value of $p$ including $p=1$.

For random symmetric networks ($\tau=1$, in Eq.\ref{Eq_mean}) with mean $p$ and variance $p(1-p)$, the 
distribution of the largest eigenvalue is given by the results in \cite{sym_rand_matrix_largest}, 
which shows that for constant $p$ the distribution of the principal eigenvalue of the random network 
is asymptotically normal with mean $Np+(1-p)$ and variance $2p(1-p)$ as $N \rightarrow \infty$ . For asymmetric 
random networks the corresponding result for the distribution of largest eigenvalue has yet to be 
found, but in \cite{asym_rand_matrix_largest} it was shown that asymmetric network of $0$ and $1$ 
entries and having mean $p$ and variance $p(1-p)$, the principle eigenvalue asymptotically converges 
to $N p$. For matrices with entries $1$ and $-1$, the mean and variance differ from above, as mean 
and variance for completely uncorrelated random network ($\tau=0$) with parameters $N,p$ would be $0$ 
and $p$ respectively.

\begin{figure}
\includegraphics[width=0.45\columnwidth,height=3.2cm]{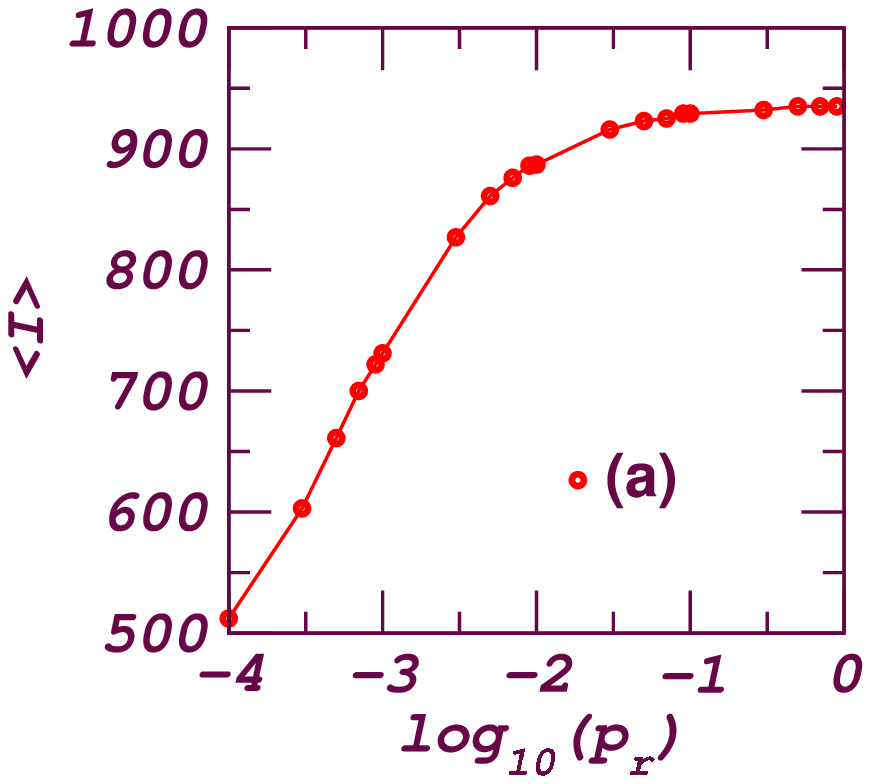}
\includegraphics[width=0.45\columnwidth,height=3.2cm]{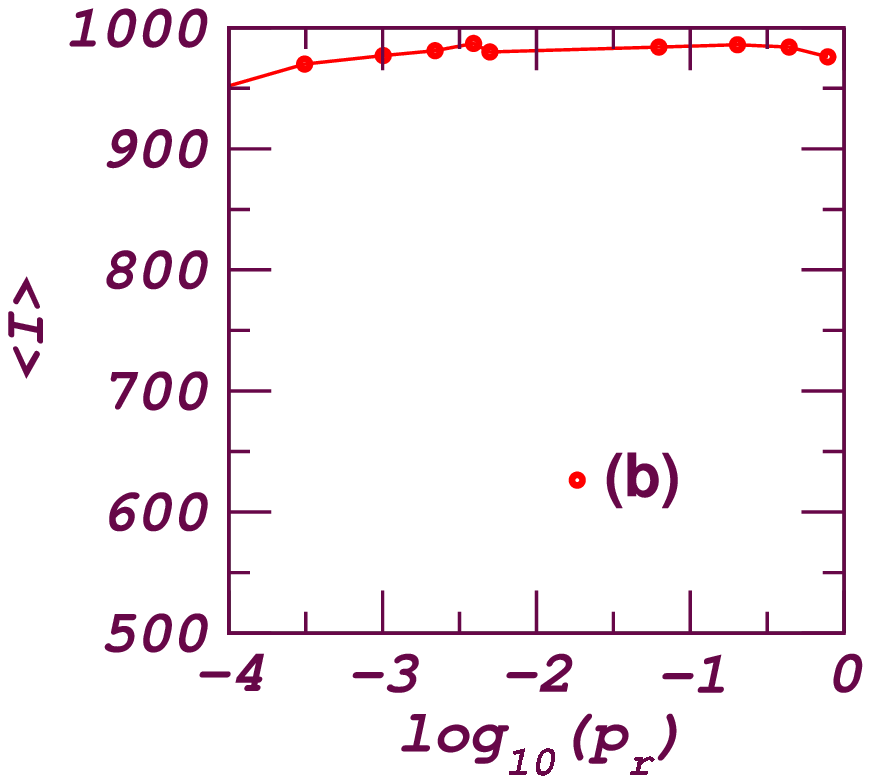}
\caption{(Color online) The IPR for (a) random network with the community as in 
Figure~\ref{fig_m_Rand} as a 
function of the
connection probability between communities $p_r$, $N=2000$ and average degree $<k>=20$
(b) a random network, $N=2000$ as in Figure~(\ref{fig_Rand}), as a function of $p$ or average degree $<k>$. }
\label{fig_m_Rand_IPR}
\end{figure}

Though the IPR is, in general, used to understand the localization property of individual eigenstates, 
the sum of all $I_i$ \\
\begin{equation}
< I > =  <\sum_i I_i>,
\label{eq_IPR}
\end{equation}
where bracket means ensemble average, can be used as a measure for localization properties of whole 
spectra altogether \cite{Seba}. We calculate the total value of 
IPR $<I>$ in order to get a measure of localization 
for the whole network. Figure~(\ref{fig_m_Rand_IPR}) plots $<I>$ for a 
random network with 
a community structure for an increasing value of the connection probability between communities. In a  
straight forward manner, the figure shows delocalization in the whole spectrum
 with deformation from the 
community structure. The value of $<I>$ increases, which means that overall
the spectrum becomes more delocalized 
untill a certain value of rewiring probability $p_r \sim 0.005$ and then 
becomes 
saturated, as shown in Figure~(\ref{fig_m_Rand}). The localization properties
the of spectra are barely distinguishable from the community structure for 
larger value of deformation. The figure is plotted 
for network size $N=2000$ and average degree $<k>=20$. We also plot the IPR (\ref{eq_IPR}) of random 
networks for various value of $p$ or an average degree $<k>$. the 
network size always remains the same, $N=2000$. 
The range for 
$p$ is from $p =0.001$ to $0.9$. We do not plot the graph below $p \sim 0.001$, since for 
$N=2000$, smaller values of $p$ yield several disconnected clusters. $N=2000$ and $p \sim 0.001$ 
yield an average degree $<k> \sim 20$.

\subsection{Tracking from correlated (symmetric) to a completely uncorrelated (asymmetric or directed) network}
\label{rand_tau}
In this section we investigate the origin of localization or delocalization of eigenstates 
by tracking the spectra as connections are made undirected. 
We start with a network having all nodes excitatory leading to a symmetric  network,
now with probability (1 - $\tau$)/2 some nodes are made inhibitory. $\tau=0$ corresponds to the case when
half of nodes are made inhibitory, leading to a completely uncorrelated network.
As $\tau$ fraction of connections are made symmetric, the mean and variance 
take form as:
\begin{eqnarray}
\mu &=& \sum_{ij} A_{ij} \sim \tau p \nonumber\\
\sigma^2 &=& \sum_{ij} (A_{ij}-\mu)^2 \sim p(1+ \tau p - 2 \tau^2 p)
\label{Eq_mean}
\end{eqnarray} 
$\tau=1$ corresponds to a symmetric network which we call
as a complete deviation from the directed network, 
i.e. $A_{i,j}=A_{j,i} \,\, \forall \, i \, \& \, j$. We call 
"deviation from directionality" because as 
one deviates from completely directed networks, spectra deviate from the circular structure. 
For a network, where nodes are inhibitory or excitatory with equal probability, $<A_{ij}>=0$ and 
$<A_{ij}^2> \sim p$. 
\begin{figure}
\includegraphics[width=0.96\columnwidth,height=5.5cm]{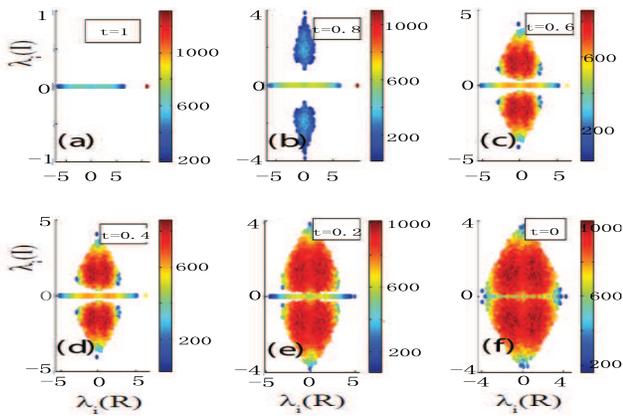}
\caption{(Color online) Spectra with the IPR for random networks having different values of $\tau$.
(a)-(f) The size and average degree of the network are $N=2000$ and $<k>=10$
respectively. For a symmetric network, $\tau=1$, and $\tau=0$ corresponds to complete
asymmetric network (see the text). The value of the IPR varies from $I=0$ 
[dark gray, blue, at the 
bottom of shading scale] to $I=1300$ [dark gray, red, at the top of the scale].
Dark gray points at both ends of the line at $x=0$ and dots at the periphery 
of the two circular region on both sides of the real axis correspond to $I=0$,
whereas a dark gray dot separated by other eigenvalues in (a) and (b), and dark
gray patches inside the circular region in (c)-(f) correspond to $I=1300$. Gray shading in
(b) denotes different levels of delocalization from $I=0$ (dark gray) to  
$I=600$ (light gray), whereas in (c) and (d), gray shading in the circular region
denotes the localization value from $I=1000$ (dark gray) to $I=600$ 
(light gray), with other gray shading for 
the intermediate values.  Light gray regions in the middle of the real axis and 
toward the top and bottom of the circular regions
in (c)-(f) correspond to the middle of the scale, whereas the darker gray region in (c) and 
(d) correspond to the value $I=800$ in the gray scale.}
\label{fig_Rand_tau}
\end{figure}

Figure~(\ref{fig_Rand_tau}) plots the spectra along with the IPR values for random networks for different 
values of $\tau$. The largest eigenvalue lies on real axis well separated from the bulk. Isolation of 
largest eigenvalue from rest is well known observation for real symmetric random matrices having non 
zero mean \cite{isolated}.It was proved in \cite{isolated} that such an isolated eigenvalue exists for 
a large matrix if the mean value of the elements is not zero and rather carries substantial fraction 
of the root mean square of the elements. Further more, the associated eigenvector would have all its 
components close to equality.

Now let us first consider the case with $\tau=1$. The corresponding matrix would be a symmetric matrix 
with $pN(N-1)$ entries being $1$ and rest being $0$, the mean and variance of this matrix can be calculated 
as $\mu=p$ and $\sigma^2=p(1-p)$ respectively (Eq.~\ref{Eq_mean}). The largest eigenvalue of the 
matrix scales as $p N$, which is equal to the average degree of the network. Rest of the eigenvalues 
are homogeneously distributed in a circular region of radius $p N(1-p)$. As connected are made directed, 
the mean $\mu$ decreases, and for $\tau=0$, mean takes 
value $0$, and variance is simply given by $p$ which is connection probability. As connections are 
made directed, eigenvalues start appearing in complex conjugate pairs, and we can divide spectra into 
two parts: (A) part of spectra with real eigenvalues, and (B) the part of spectra having complex 
conjugate pairs of eigenvalues. For $\tau =0.8$, which means $20\%$ of connections are made directed, 
many eigenvalues are still in group (A), i.e., they lie on real axis. Complex conjugate 
pairs of eigenvalues form an oval shape that is well separated from real axis (see 
Figure~(\ref{fig_Rand_tau}). The localization properties of eigenstates from both the groups do not 
change much, as bulk part which is separated from the real axis is still delocalized, only middle part of 
eigenstates lying at the real axis (group (A)) delocalized than a completely symmetric network. 
As more connections are made 
directed, complex conjugate pairs of eigenvalues start appearing with delocalized states. Bulk of the
complex conjugates pairs lying at both sides of the real axis and forming oval shape 
(group (B)) are delocalized, 
only few eigenvalues in this group with largest absolute value remain localized. Eigenvalues forming 
group (A) also show similar features as group (B), eigenstates with larger eigenvalues at both ends 
remain localized, whereas rest of the eigenstates are more delocalized than earlier.

As more connections are made directed (value of $\tau$ is decreased), size of group (B) keeps 
increasing, and consequently two parts of eigenstates come closer. For $\tau=0.2$, group (B) is no 
longer separated from the eigenstates lying on the real axis (group (A)).

\begin{figure}
\includegraphics[width=0.9\columnwidth,height=3cm]{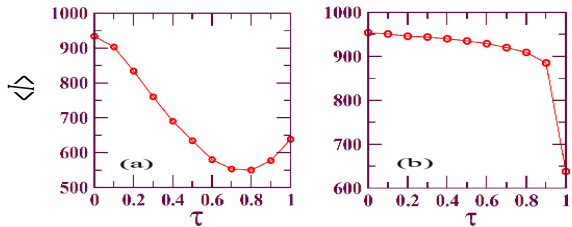}
\caption{(Color online) The total IPR as a function of $\tau$. The data are
plotted for $N=2000$, $<k>=20$
and for 10 ensemble averages of the random networks. (a) Networks with inhibitory and
excitatory nodes leading to $1$ and $-1$ entries in the corresponding matrix, and (b)
random network with only excitatory nodes leading to only $1$ entries in the 
corresponding matrix.}
\label{fig_Rand_Total_IPR_Tau}
\end{figure}

Figure~(\ref{fig_Rand_Total_IPR_Tau}) plots the total $<I>$ (Eq.\ref{eq_IPR}) for a particular
value of connection probability $p=0.01$ and network size $N=2000$. Starting from a symmetric network
($\tau=1$), connection are made directed with probability $(1-\tau)$. 
Overall spectral delocalization increases as connections are made
directed leading to occurrence of complex conjugate pairs of eigenvalues in spectra.
Delocalization reaches its maximum for $\tau=0$ value which corresponds to a complete uncorrelated or 
directed network. For networks having only positive entries, there is not much changes in 
the localization properties till value $\tau \le 0.8$, after this there is a sharp decay in 
$<I>$ indicating sudden localization transition. For values between $0$ and $0.8$, spectral
distribution smoothly changes from circular to elliptical, localization property does not show any
rich structure, as shown for the previous case.
 
\subsection{Random networks with positive entries}
In the above we investigated networks with both excitatory and inhibitory connections. This section
considers all nodes being excitatory, and consequently 
the adjacency matrix has all positive entries only. 
For asymmetric random network of $0$ and $1$
entries, with mean $p$ and variance $p(1-p)$, the principle eigenvalue asymptotically converges
to $N p$ \cite{asym_rand_matrix_largest}. The rest of the eigenvalues are distributed in
circular region with radius given by $\sqrt{N p (1-p)}$.
In this section we present the spectral changes as
network goes from an asymmetric (directed) corresponding to $\tau=0$, to a symmetric network 
corresponding to $\tau=1$.
The mean and variance for networks with entries $1$
would not depend on the value of $\tau$, and depends only on connection
probability $p$ as\\
\begin{eqnarray}
\mu = p \nonumber\\
\sigma^2 = (1-p) 
\end{eqnarray}
\begin{figure}
\includegraphics[width=0.9\columnwidth,height=5.8cm]{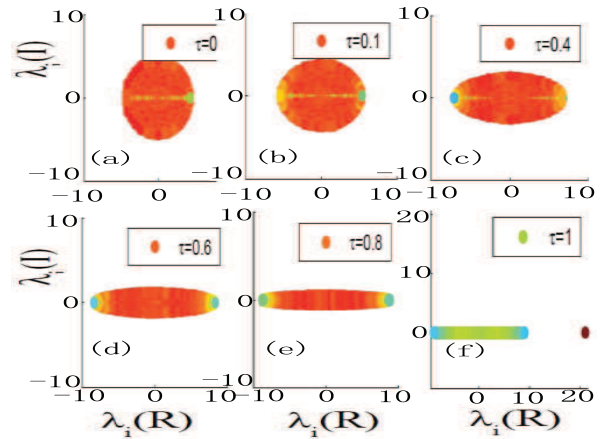}
\caption{(Color online) Random network for various values of $\tau$ but with only excitatory coupling. The network
size $N=2000$ and the average degree $<k>=20$. The value of the IPR varies from $I=0$ [light gray, blue, at the two ends of the spectra on the x axis] to $I=1200$ [dark gray, red, circular regions]. the light gray
region in the middle of x-axis in (f) is not as delocalized as the light gray 
points toward both ends. The light gray horizontal line at x axis in (a)-(c), 
and the light gray patches at the corner of (b)-(e) have
the same level of localization, which shows less delocalization than the light gray region
in (f) described above.}
\label{Only1_Rand_tau}
\end{figure}

Figure~(\ref{Only1_Rand_tau}) shows spectra along with localization property for different values 
of $\tau$. Spectral distribution smoothly goes from circular distribution to
elliptical as value of $\tau$ increases. For all values
of $\tau$, the largest eigenvalue is completely delocalized, and situated very well separated from 
the rest bulk part with value $\lambda_{max} \sim N p$, as for the case $\tau=0$. Note that
for each value of $\tau$, the expected number of entries $1$ remains same for a fixed $p$ value.

\section{Networks with scalefree community structure}
\label{m_scale}
In this section we consider that each community is modeled by a scalefree network instead of
a random network. Each sub-network has size $N=1000$ and
average degree $<k>=20$, and is constructed using BA preferential attachment rule \cite{BA}. 
With probability $p_r$ intra-community connections are rewired into inter-community connections.
Connections are made directed in same manner as given in the section \ref{m_Rand}.
\begin{figure}
\includegraphics[width=0.9\columnwidth,height=5.5cm]{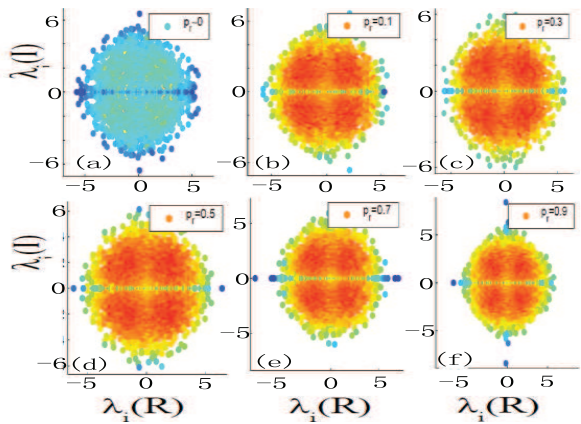}
\caption{(Color online) Eigenvalues with the IPR for networks having two scalefree sub-networks, and 
for different rewiring probabilities. The size of
each community $N=1000$ and the number of communities is $m=2$. The average 
degree is $<k>=20$.
The rewiring probability is denoted by $p_r$. The value of IPR varies from $I=0$ [dark gray, blue, scattered points at the peripheral of all sub-figures and a 
few points separate from
the bulk in (d)-(f)] to $I=1200$ [dark gray, red, patches in the middle of 
(b)-(f)]. Gray shading in (a) shows different levels of delocalization with dark gray being
the most delocalized to lighter gray being less delocalized. Light gray points 
at the periphery in (b)-(f) are not as delocalized as the light gray region inside the circle of (a), whereas they are more delocalized than the light gray region inside the circle in
(b)-(f). }
\label{fig_m_Scale}
\end{figure}

Spectra of the above construction are plotted in Figure(~\ref{fig_m_Scale}). The first difference between 
the spectra of these networks from those having a random community structure 
(\ref{m_Rand}) is that 
the eigenvalues are not homogeneously distributed in the circular region. The spectra are more dense 
around origin, and sparser near boundary. Also the value of largest absolute eigenvalue increases, as 
the highest degree of the node in the current structure are very much larger than that of the previous 
structure (\ref{m_Rand}). Localization property overall remains same as those of the networks 
discussed in \ref{m_Rand}.

\begin{figure}
\includegraphics[width=0.93\columnwidth,height=7.4cm]{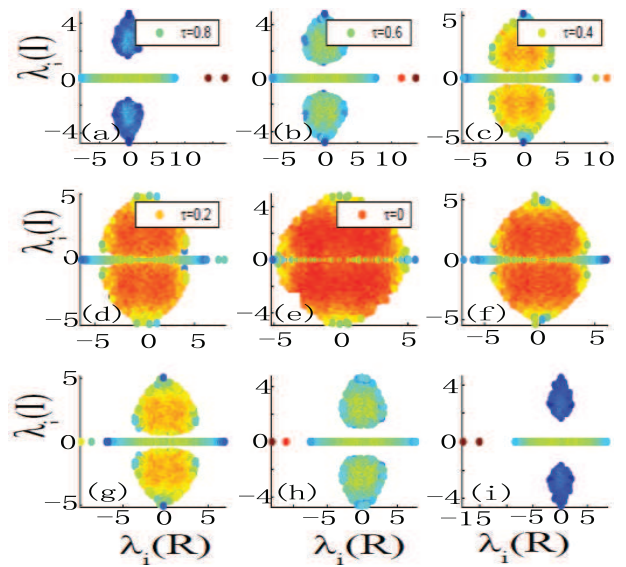}
\caption{(Color online) Spectra with the IPR for networks having two random sub-networks, and 
for $p_r=0.2$ rewiring 
probability. The size of each community $N=1000$ and the average degree 
$<k>=20$. The different probabilities of directed connections are plotted
(see the text). Configurations are shown in which (a)-(d) excitatory nodes are 
dominant, (f)-(i)  inhibitory nodes are dominant, and (e) there is
equal probability of a node being excitatory or inhibitory.
The value of the IPR varies from $I=0$ [dark gray, blue, patches at the 
periphery of both  circular regions on two sides of the real axis in (a) and 
(i), and dark gray points toward both ends of the real x axis in (c)-(g)] to 
$I=1200$ [dark gray, red, points well separated from bulk in (a)-(i)]. Gray shading in (a),(b), (h) and (i)
 denotes different levels of
delocalization with dark gray (except two dots well separated from the bulk
) denoting maximum and light gray showing minimum delocalization. Light gray area inside the
circular regions in (c) and (g) show less delocalization than the light gray area in 
(a) and (b). Dark gray area inside the circular regions in (c) - (g) show less delocalization than
the similar dark gray area inside the circular region of (a).} 
\label{fig_m_Rand_Tau}
\end{figure}

\section{Community structure : tracking from symmetric to a complete 
asymmetric (directed) network}
\label{m_Community_Tau}
In this section we investigate the origin of localization or delocalization properties of eigenstates 
for networks having distinguished community structure. Particularly we concentrate on the community 
dependent 
localization features as connections are made directed, and compare them with random networks 
features presented in the 
section~\ref{rand_tau}. Starting with $m=2$ random symmetric sub-networks with $p_r=0.2$ rewiring 
probability between these sub-networks, connections are made directed with the probability 
$\tau$ as explained in the previous
section. Again, $\tau=0$ corresponds to a 
fully directed network such as explained in \ref{m_Rand}, and $\tau=1$ corresponds to 
the symmetric network. A symmetric random network with connection probability $p$ 
has mean $p$ and variance $p(1-p)$, whereas mean and variance of a fully directed network (made of 
$1$ and $-1$) are $0$ and $p$ respectively. Probability of directed connections
($\tau$) affects 
the mean and variance as given in Eq.~\ref{Eq_mean}.

Figure~(\ref{fig_m_Rand_Tau}) plots the eigenvalue spectra combined with the IPR as $\tau$ varies 
from $\tau=1$ (a perfect symmetric case) to $\tau=0$ (a uncorrelated case).  Subfigure (a)
corresponds to the configuration when almost $80\%$ nodes are excitatory leading to very large 
correlation among entries of the matrix.  For very small number of 
directed connections the spectra have mostly real eigenvalues, and eigenstates are localized. As 
number of directed connections increases spectra have more complex eigenvalues. Subfigure (i)
corresponds to the case where $\sim 80\%$ of nodes are inhibitory, which is exactly the same case
in terms of correlations as shown for subfigure (a) except that the corresponding network has dominating
$-1$ entries.

As similar to the section \ref{rand_tau}, spectra can be divided into two parts, (A) part of spectra 
lying on the real axis, and (B) the part which form complex conjugate pairs. The appearance of complex 
eigenvalue seems to be an indicator of delocalization transition. For small number of directed connections 
(Figure~\ref{fig_m_Rand_Tau} a) most of the eigenvalues are real and hence localized. As 
more connections are made directed, complex pairs of eigenvalues start appearing and eigenstates associated with 
these complex values start getting delocalized. For $\tau=0$, which leads
to a complete asymmetric or directed network, one gets 
complex pairs of eigenvalues covering the whole circle, and hence almost all eigenstates, except those 
associated with real eigenvalues and those with maximum absolute value, get delocalized.

Now, as we compare the spectra of above construction with those of random networks, we observe one 
remarkable difference revealing two communities hidden in the network. For small 
number of directed connections  
(i.e. large values of correlation parameter $\tau$), there are exactly {\it two} isolated eigenvalues 
on the real axis well separated from the bulk. Figure~(\ref{fig_m_Rand_Tau} a) shows that 
these isolated eigenstates corresponds to two largest eigenvalues and both of them are delocalized. 
For $\tau=0.6$, which means that number of directed connections is higher 
than $\tau=0.8$, these isolated 
eigenvalues still exist on the real axis but come closer to the rest bulk part. Further increase 
in the number of directed connections  brings isolated eigenvalues even closer, as well their localization 
properties get affected. For $\tau=0.4$, these isolated eigenvalues are no more extended, and 
eigenstates having largest absolute values from group (A) and (B) are most localized. As $\tau$ is 
decreased further, these isolated eigenvalues merges with the bulk part of group (A).

\begin{figure}
\includegraphics[width=0.9\columnwidth,height=3.6cm]{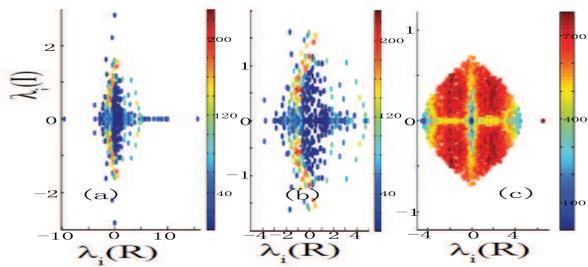}
\caption{(Color online) (a) Spectra of the zebrafish metabolic network. (b) 
Close-up of the spectra. (c) Spectra of the random network with same parameters (i.e. number of nodes,
and number of directed and undirected edges) as zebrafish metabolic network. Gray shading (color)
denotes the localization properties of the corresponding eigenvalues. 
Dark gray (blue) patches in 
(a)-(b) and dark gray dots at the end of the real axis as well at the center of (c) correspond to the bottom of the gray shading with an IPR value $I=0$, 
whereas dark gray (red) patches 
inside the circular region in (c) and a dark gray dot well separated from bulk lying at extreme right correspond to the top of the shading scale with an 
IPR value $I=800$. 
The light gray vertical regions toward the left side in (a) and (b) is
not as  delocalized (with an IPR value close to 500, as in the scale) as 
the light gray vertical region toward the right side of the same plot (with an 
IPR value varying by about 300 
in the scale). Similarly, light gray points 
at the right corner of (c) are more 
delocalized than the part of the vertical and horizontal gray area of similar shade crossing at center. Darker gray patches in (c) are not as delocalized 
as  the patches with the same
shade in (a) and (b).} 
\label{fig_Zebrafish_Metabolic}
\end{figure}

\section{Zebrafish metabolic network}
This section compares the behavior shown by the model networks discussed so far with
a realworld network. We use network for the zebrafish 
metabolic system. The network is constructed by using the KEGG data base \cite{kegg}. Nodes in 
the network are reactions. If the product of one reaction (say A) is the 
substrate of another reaction  
(say B) then there is a 
directed link from A to B. There are also some directed links that correspond to the 
irreversible biochemical reactions. The largest connected component of the 
network has $N=1593$ nodes and 
$E=9018$ edges. The number of undirected edges is $6770$ and the number of directed edges is $2248$. The value of 
$\tau$ can be calculated as $\tau=6670/9018 \sim 0.7$ The average degree of the network is $E/N \sim 
6$. Figure~(\ref{fig_Zebrafish_Metabolic}) plots the spectra of this network. In order to compare the 
results with a corresponding random network, we plot 
Figure~(\ref{fig_Zebrafish_Metabolic} c)  
spectra of a random networks ensemble with the same size and number of directed and 
undirected edges as in the 
zebrafish network.

We notice that the spectrum of the zebrafish network has few drastic 
differences from those of model networks 
discussed in the preceding sections. The main observation is the clustering of eigenvalues 
in an elliptical disk along the imaginary axis. Two extremal eigenvalues lie on the real axis, and they 
are far from the rest of the bulk eigenvalues. The corresponding random network also has the largest 
eigenvalue situated at the real axis with a value close to the average degree of the network. This 
eigenvalue is situated far from the rest of the bulk eigenvalues centered around zero. This is a typical 
spectral behavior of directed random networks, as discussed in the 
Sec.~\ref{Rand}.

Furthermore, the localization properties of eigenstates of the real networks studied show 
very different phenomena than what observed for the corresponding model networks. In an elliptical disk 
there is one layer of eigenstates that are very much localized, and the second layer in the disk shows 
a full range of IPR values between extreme localization and extreme delocalization. There are many 
eigenvalues that lie on the real axis, which can be explained by the relation between asymmetry in the matrix 
and the occurrence of complex eigenvalues discussed in Sec. VI 
(Fig.~\ref{fig_m_Rand_Tau}). However the localization properties of these eigenstates corresponding 
to the real axis are different than the corresponding model network and other earlier model networks we 
investigated in the preceding sections. The main difference is that all eigenstates lie toward the
bottom on the localization scale. The isolated eigenvalues for model networks so far are
observed with the most extended state, whereas for the zebrafish network, the isolated eigenvalue lies 
toward bottom of the localization value.

\section{Conclusion and Discussions}
We analyzed the spectral properties of various model directed networks and a realworld network.
The network edges take value $1$ and $-1$ depending upon whether it is starting from
an excitatory node or from an inhibitory node. 
If all nodes are excitatory
then the corresponding network is symmetric ($\tau=1$). Directionality is introduced by making some
nodes inhibitory, and consequently the corresponding edges take value $-1$.
An equal expected value of inhibitory and excitatory nodes
gives rise to a completely uncorrelated network ($\tau=0$).

Asymmetric networks ($\tau=0$) with a community structure modeled by a 
random sub-network
show circular spectra 
similar to random uncorrelated matrices. Spectra for networks with 
entries $1$, $-1$ follow Girko's circular
distribution with radius $\sqrt{pN(1-p)}$. For a very well distinguished
community structure $p_r \sim 0$, the whole spectrum is localized except
for some of eigenstates at the boundary. As $p_r$ increases and the network
deviates from a community structure there is a sudden change in the localization
property for a very small value of deformation from the perfect community
structure. The value as less as $p_r \sim 0.1$, which corresponds to only
$10\%$ inter-community connections (Fig.~\ref{fig_m_Rand}), gives rise to a
delocalized spectrum. In order
to understand this delocalization transition we track the spectra for very
fine rewiring between communities (Fig.~\ref{fig_m_Rand}). For 
exactly one rewiring most of the eigenvalues lying on the real axis
are delocalized. This suggests that there are eigenfunctions confined within their
respective random communities. With the increase in rewiring between communities
some of the eigenstates lying at the boundary of the 
circular
region start delocalizing. This suggests that some of the eigenstates now span
the whole network through rewired connections that connect two sub-networks. 
With the further deformation from 
the community structure, the next level of delocalization starts occurring through a
circular ring around center, which gradually spans the whole spectrum with an 
increase in the  rewiring.

The spectra of random networks, where the probability for a node being inhibitory or 
excitatory is equal, show a circular distribution with a radius  
$\sqrt{pN(1-p)}$. Based on the IPR values, the spectrum can
be divided into two parts, (A) one part consisting of eigenstates at the real axis and at four corners with large absolute
eigenvalues that are localized, and (B) another part consisting of 
the bulk middle part of the spectrum, which is less localized. 
As the connection probability
increases part (B) starts dominating the spectra, except for a 
few localized eigenvalues, which remain very well
separated from the bulk part even for very large connection probabilities.

In order to understand the mechanism for localization we tracked the 
spectra as the network is rewired from a completely symmetric structure ($\tau=1$) to
a completely asymmetric one ($\tau=0$). For symmetric networks spectra lie 
on the real axis with exactly one eigenvalue separated from the bulk. As 
connections are made directed by making some nodes inhibitory, some of the 
eigenstates
start occurring in complex conjugate pairs. The eigenvalue distribution along with the IPR value
show a rich pattern. Overall, the spectra gradually become 
more delocalized as the number of directed connections increases. 

Furthermore we studied spectra of matrices with non-negative entries. This
kind of matrices was previously investigated in details; we also showed that
the spectral distribution follows Girko's generalized law. For $\tau=0$ it
shows circular distribution, and as the correlation $<A_{ij}A_{ji}>$ increases
spectra become elliptical. The localization properties of whole spectra remain
similar except for a sudden localization for very large values of $\tau$. 

Spectra of networks with a scalefree community structure do not show
a localization pattern that is much different from spectra of networks with 
the random community structures. 
For two separated communities with one connection between them, the
spectra are circular with radius $\sqrt{pN(1-p)}$ and the boundary of the circular
region is scattered. As more connections are rewired leading to stronger coupling between
the communities,
the spectral distribution remains same but the localization behavior of the
eigenstates
changes. The main difference from the spectra of 
networks in section Sec.~\ref{m_Rand} is that they are less delocalized which is
obvious because any network with a scalefree community structure
always has more structure than 
the corresponding networks
with a random community structure. The crucial point 
 is that the spectral
distribution is very robust with the changes in the structure of the network
and depends strongly on the directionality measure $\tau$, whereas the 
localization
property depends on both the structure of the networks and the  
directionality of connections, i.e., the value of $\tau$.

We encounter interesting features while tracking the spectra as the network with
a community structure goes from a directed to a symmetric one. For a
random network with two random communities, the spectrum shows exactly
two eigenvalues
well separated from the bulk (Fig.~\ref{fig_m_Rand}). 
These eigenstates
correspond to the two largest absolute values and are delocalized. The largest one
is completely delocalized as was the case for a random network
without a community structure, but the second largest one that is well 
separated from the 
bulk provides a distinguished signature of two communities in the network. 
As the number of directed connections increases the spectrum can again
be divided into two parts, (A) the part of the spectrum with real eigenvalues
and (B) the part of the spectrum with complex conjugate pairs.

To the end we analyze a realworld spectrum. Spectral properties of the metabolic
network
of zebrafish show much richer features than the model networks
considered in the present paper. One very striking difference is the localization
property of the eigenstate with the largest absolute value.  This eigenstate is localized
in contrast to those of the model networks. Another observation is the clustering of eigenvalues in an
elliptical disk along the imaginary axis. 

In the paper we have investigated spectra and localization properties
of directed networks with binary entries. The networks with inhibitory
and excitatory nodes have much richer spectra than the networks with only
excitatory nodes. The bulk of the spectra for completely asymmetric 
networks follow Girko's law, but as probability of directed connections 
is reduced the spectra
show very different patterns depending upon the network structure
and the ratio of inhibitory and excitatory nodes. The realworld network
that we studied here shows a very different spectral pattern from any
of the model networks we have studied. Though directed networks span a 
variety
of complex systems, the research for directed networks leading to complex eigenvalues is limited.
The results presented in the present paper provide
a useful platform to understand the structural pattern in 
directed networks, and can be used to investigate further the dynamical
behavior of nodes relevant to a variety of problems ranging from physics to sociology.

\section{Acknowledgment}
We acknowledge Dr. Changsong Zhou for useful discussions at the initial phase of the work, and 
Zhang Xun for providing zebrafish metabolic data from KEGG \cite{kegg}.

\end{document}